\documentclass[nanomaterials,article,submit,pdftex,moreauthors]{mdpi} 
\firstpage{1} 
\pubvolume{1}
\issuenum{1}
\articlenumber{0}
\pubyear{2023}
\copyrightyear{2023}
\datereceived{ } 
\daterevised{ } 
\dateaccepted{ } 
\datepublished{ } 
\hreflink{https://doi.org/} 

\renewcommand{\d}{\mathrm d}
\usepackage{bm}
\usepackage{braket}
\newcommand{\addDima}[1]{#1}

\Title{Optical orientation of excitons in a longitudinal magnetic field\\ in indirect band gap (In,Al)As/AlAs quantum dots with type-I band alignment}

\TitleCitation{Title}

\Author{T.~S.~Shamirzaev$^{1}$*\orcidA{}, A.~V.~Shumilin$^{2}$\orcidB{}, D.~S.~Smirnov$^{2}$*\orcidC{}, D.~Kudlacik$^{3}$\orcidF{}, S.~V.~Nekrasov$^2$\orcidH{}, Yu.~G.~Kusrayev$^2$\orcidG{}, D.~R.~Yakovlev$^{3,2}$\orcidD{} and M.~Bayer$^{3}$\orcidE{}}

\AuthorNames{Firstname Lastname, Firstname Lastname and Firstname Lastname}

\AuthorCitation{Lastname, F.; Lastname, F.; Lastname, F.}

\address{%
$^{1}$ \quad Rzhanov Institute of Semiconductor Physics, Siberian Branch of the Russian Academy of Sciences, 630090 Novosibirsk, Russia\\
$^{2}$ \quad Ioffe Institute, Russian Academy of Sciences, 194021 St.Petersburg, Russia\\
$^{3}$ \quad Experimentelle Physik 2, Technische Universit\"at Dortmund, 44227 Dortmund, Germany
}

\corres{Correspondence: sha\_tim@mail.ru and smirnov@mail.ioffe.ru}

\abstract{The exciton recombination and spin dynamics in (In,Al)As/AlAs quantum dots (QDs) with indirect band gap and type-I
band alignment are studied. The negligible (less than 0.2~$\mu$eV) value of the anisotropic exchange interaction in these QDs prevents
a mixing of the excitonic basis states with pure spin and allows for the formation of spin polarized bright excitons for quasi-resonant circularly
polarized excitation. In a longitudinal magnetic field, the recombination and spin dynamics of the excitons are controlled by the
hyperfine interaction between the electron and nuclear spins. A QD blockade by dark excitons is observed in magnetic field eliminating the impact of the nuclear spin fluctuations. A kinetic equation model, which accounts for the population dynamics of the bright
and dark exciton states as well as for the spin dynamics, has been developed, which allows for a quantitative description of the experimental data.}


\begin{document}


\section{Introduction}
\label{sec:intro}

Spin-dependent phenomena in semiconductor heterostructures are
attractive from the viewpoints of both basic
physics~\cite{Dyakonov,Glazov} and potential
applications~\cite{Fert,Bader}. Semiconductor quantum dots (QDs) are
of great interest as objects with long spin lifetimes of electrons
and holes, as the key obstacle for spin-based quantum information
processing is spin relaxation. Indeed, the carrier localization
slows down the spin relaxation due to suppression of the mechanisms
determining the relaxation of freely moving charge
carriers~\cite{Khaetskii0}. Therefore, the spin relaxation time of
electrons localized in QDs can reach milliseconds, as confirmed
experimentally~\cite{Kroutvar}.

A common approach to study the spin dynamics is optical
orientation provided by circularly polarized light~\cite{OO_book}.
The light delivers angular momentum to the electron spin system, inducing
its polarization, which subsequently decays due to
relaxation processes. The spin dynamics can be measured by the decay
of the photoluminescence (PL) circular polarization
degree~\cite{OO_book}. However, this technique is not suitable to
study the spin dynamics of excitons in direct band gap QDs at zero
magnetic field. The axial symmetry breaking, which always occurs in
experimentally available QDs, leads to a mixing of the bright pure spin exciton states via
the anisotropic exchange interaction~\cite{Glazov,Rautert100}.
Therefore, until recently, optical orientation at zero magnetic
field has been used in QDs mainly to study charged excitons (trions)
formed, for example, from a pair of electrons and a
hole~\cite{Taylor}. In this case, the ground state of the trions is an electron spin-singlet for
which the exchange interaction with the hole vanishes~\cite{Dunker2012,Shamirzaev106}.

We have recently demonstrated the suppression of the
anisotropic exchange interaction in indirect band gap
(In,Al)As/AlAs QDs, which prevents the bright exciton
mixing~\cite{Rautert99}. Additionally, the weak electron-nuclei
interaction in the X valley makes the electrons in such QDs
relatively robust against spin decoherence~\cite{Kuznetsova101}.
These features of the exchange and hyperfine interactions allowed
for the discovery of the dynamic electron spin polarization effect,
which takes place for unpolarized optical excitation in magnetic
fields of the order of a few
millitesla~\cite{Smirnov125,Shamirzaev104a}. This effect was later
described also for organic semiconductors~\cite{Organic} and
moir\'e QDs~\cite{Moire}, where it was recently observed
experimentally~\cite{Exp_moire}.

In this paper, the exciton recombination and spin dynamics in
(In,Al)As/AlAs QDs with indirect band gap and type-I band alignment
are studied in a longitudinal  magnetic field under optical
orientation. The magnetic fields are moderately weak, so that all
Zeeman splittings of the spin states are much smaller than the
thermal energy. These experimental conditions prevent a circular
polarization of the exciton emission to be induced by the magnetic
field~\cite{Ivchenko60,Shamirzaev94,Shamirzaev96,Shamirzaev104,Shamirzaev102}.
Measuring the optical orientation in (In,Al)As/AlAs QDs with
modulation of the sign of the circular polarization of the exciting
light reveals a dependence of the PL circular polarization degree
on the modulation frequency, which arises due to the long times of
exciton recombination and spin relaxation in indirect gap QDs.
Different protocols of the spin orientation measurement are
compared, where the effect of QD blockade by dark
excitons is found.

The paper is organized as follows. In Sec.~\ref{sec:1} the studied
heterostructures and used experimental techniques are described. In
Sec.~\ref{sec:2} we present the experimental data including
time-resolved unpolarized PL, PL under selective excitation in zero
magnetic field, the recovery of PL circular polarization in
longitudinal magnetic fields for continuous-wave  ($\emph{cw}$)
circularly polarized excitation, and the effects of different
excitation-detection  protocols. Then, in Sec.~\ref{sec:4} the theory
of exciton spin dynamics in QDs is presented and compared with the
experiment.

\section{Experimental details}
\label{sec:1}

The studied self-assembled (In,Al)As QDs embedded in an AlAs matrix
were grown by molecular-beam epitaxy on a semi-insulating
$(001)$-oriented GaAs substrate with an 400-nm-thick GaAs buffer
layer~\cite{Shamirzaev78}. The structure contains 20 layers of
undoped (In,Al)As/AlAs QDs sandwiched between 25-nm-thick AlAs
layers. The nominal amount of deposited  InAs is about $2.5$
monolayers. Lens-shaped QDs with an average diameter of $15$~nm and a
height of $4$~nm were formed at the temperature  of
$520^{\circ}$C with the growth interruption time of 20 s. The QD density
is about $3 \times 10^{10}$~cm$^{-2}$ in each
layer. A 20-nm-thick GaAs cap layer protects the top AlAs barrier
against oxidation. Further growth details are given in
Ref.~\cite{Shamirzaev78}. The interlayer distance and QD density
were chosen  to prevent an electronic coupling between individual
quantum dots~\cite{Shamirzaev2010,ShamirzaevSST}. The growth axis
$z$ coincides with the (001) crystallographic direction. Note, that
the band gap energy of the GaAs substrate is 1.52~eV and that of the
AlAs barrier is 2.30~eV~\cite{Vurgaftman}.

The sample was mounted strain free inside a cryostat with a variable
temperature insert. The temperature was varied from $T=1.7$~K up to
20~K.  Magnetic  fields in the mT range were generated by an
electromagnet with an accuracy better than 0.1~mT. The magnetic field direction coincided with
the structure growth axis ($z$), along which also the wave vector of the excitation light was pointing
(Faraday geometry).

The PL was excited either non-resonantly with the photon energy of a
laser exceeding considerably the emission energies in the QD
ensemble, or selectively with a laser energy tuned to a value within the
inhomogeneously broadened exciton emission band of the QDs. The
nonresonant excitation was provided by the third harmonic of a
Q-switched Nd:YVO$_4$ pulsed laser with the photon energy of
3.49~eV, the pulse duration of 5~ns and the repetition rate of $2$~kHz.
The excitation density was kept below
100~nJ/cm$^2$~\cite{Shamirzaev84}. For selective excitation a
$\emph{cw}$  Ti:Sapphire laser with a photon energy tunable in the
spectral range from 1.50 to 1.75~eV was used.

For the time-resolved and time-integrated PL measurements we used a
gated charge-coupled-device camera synchronized with the laser via
an external trigger signal. The time between the pump pulse and the
start of the PL recording, $t_{\text{delay}}$, could be varied from
zero up to $1$~ms. The duration of PL recording, i.e. the gate
window $t_{\text{gate}}$, could be varied from $1$~ns to
$500~\mu$s. The signal intensity and the time resolution of the
setup depend on $t_{\text{delay}}$ and $t_{\text{gate}}$. The
highest time resolution of the detection system is $1$~ns.

\begin{figure}[t]
\includegraphics[width=\textwidth]{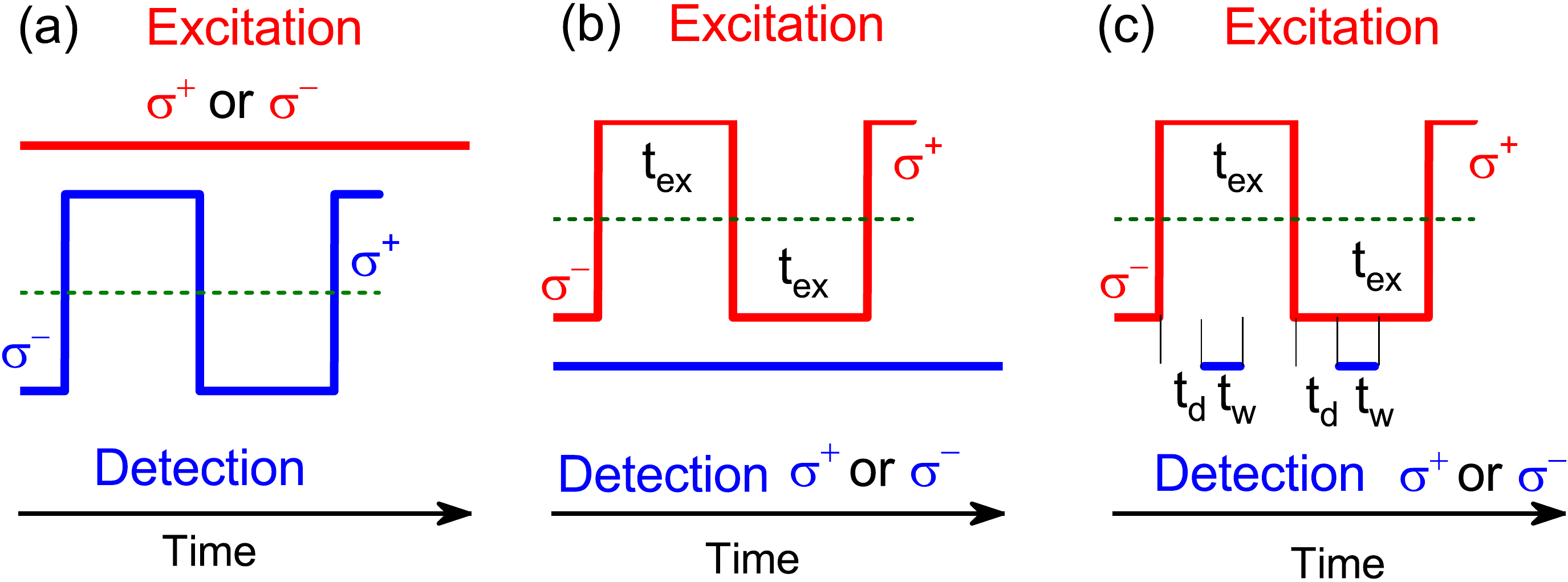}
\caption{\label{fig1} Protocols used for measurement of
$\rho_c$: (a) $\emph{cw}$ circularly ($\sigma^+$ or
$\sigma^-$) polarized excitation and modulation of polarization in the
detection channel. (b) $\emph{cw}$  measurement of the emission in
$\sigma^+$ or $\sigma^-$ polarization and modulation of
polarization in the excitation channel with the period 2$t_{ex}$.
(c) Modulation of polarization in the excitation channel with
the period 2$t_{ex}$ and measurement in the time window $t_{w}$ with
delay $t_{d}$ after changing the excitation polarization.}
\end{figure}

For measuring the optical orientation (optical alignment) effects,
the circular (linear) polarization of the excitation laser and of the PL emission
were selected by corresponding combination of circular (linear) polarizers (Glan-Thompson prism)
as well as quarter-wave (half-wave) plates. For the optical
orientation measurement, the circular polarization degree of the PL
induced by circularly-polarized excitation, $\rho_\text{c}$, is
recorded:
\begin{align}
\rho_\text{c}=\frac{I^{+/+} - I^{+/-}}{I^{+/+} + I^{+/-}}.
\end{align}
Here $I^{\text{a/b}}$ is the intensity of the
$\sigma^\text{b}$-polarized PL component measured for
$\sigma^\text{a}$-polarized excitation. The labels $+$ and $-$
correspond to right-hand and left-hand circular polarization,
respectively. Note that in our experiments the circular
polarization degree induced by the external magnetic field is
negligible, since the Zeeman splitting of the excitonic states in the applied
magnetic field strength range is much smaller then the thermal energy.

In the optical alignment measurement, the linear polarization degree
of the $\emph{cw}$  PL ($\rho_\text{l}$) induced by linearly
polarized excitation is measured. The linear polarization degree is
defined as
\begin{align}
\rho_\text{l}=\frac{I^\text{0/0}-I^\text{0/90}}{I^\text{0/0} +
I^\text{0/90}},
\end{align}
where $I^{\text{a/b}}$ are the PL intensities with the superscripts
$a/b$ corresponding to the direction of the excitation/detection linear
polarization. The direction ``0'' is parallel to the [110]
crystallographic direction and the direction ``90'' is parallel to
the $[1\bar 10]$ direction.

The electron spin dynamics was investigated by measuring the PL
polarization degree for optical orientation in longitudinal magnetic
fields. In these experiments, the PL was detected by a GaAs
photomultiplier combined with a time-correlated photon-counting
module. Three protocols were  used: (i) $\emph{cw}$ circularly
polarized excitation ($\sigma^+$ or $\sigma^-$ ) and measurement of
$\rho_c$ using an acousto-optic quarter-wave modulator with
modulation frequency $f_m= \frac{1}{2\times t_{\text{ex}}}=50$~kHz,
where ${2\times t_{\text{ex}}}$ is the modulation period
[Fig.~\ref{fig1}(a)]; (ii) modulation of excitation polarization
via an electro-optic half-wave modulator before the quarter-wave
plate (with $f_m$ in the range from 1 up to 500~kHz) and $\emph{cw}$
measurement of the emission in $\sigma^+$ or $\sigma^-$ polarization
[Fig.~\ref{fig1}(b)]; and (iii) the same excitation scenario as in
protocol (ii) but measurement with a delay $t_{\text{d}}$
after changing the excitation polarization (from $\sigma^+$ to
$\sigma^-$ and vice versa) during the time window $t_{\text{w}}$
[Fig.~\ref{fig1}(c)].

\section{Experimental results}
\label{sec:2}

The dispersion of  the  QD size, shape, and composition within
the ensemble leads to  the  formation of (In,Al)As/AlAs QDs with
different band structures~\cite{Shamirzaev78}, as shown in
Fig.~\ref{fig2}(a). The electron ground state changes from the
$\Gamma$ to the X valley for decreasing dot diameter, while the
heavy-hole (hh) ground state remains at the $\Gamma$ point, see
Fig.~\ref{fig3}(a). This corresponds to a change from a direct to
an indirect band gap in momentum space, while the type-I band
alignment is preserved, that is, in both cases, electron and hole
are spatially confined within the (In,Al)As
QDs~\cite{Shamirzaev78,Shamirzaev84,ShamirzaevAPL92}.

Recently, we demonstrated that the coexistence of
(In,Al)As/AlAs QDs with direct and indirect band gaps within an
ensemble results in a spectral dependence of the exciton
recombination times. In the momentum-direct QDs, the excitons
recombine within a few nanoseconds. On the contrary, the
momentum-indirect QDs are characterized by long decay times due to
the small exciton oscillator
strength~\cite{Shamirzaev78,Shamirzaev84,ShamirzaevAPL92,Abramkin112,Shamirzaev60,Shamirzaev95,Abramkin103,
Shamirzaev97}. Here, we use time-resolved PL to select the indirect band gap QDs.

\subsection{Time-resolved unpolarized PL}

\begin{figure}[]
\includegraphics[width=\textwidth]{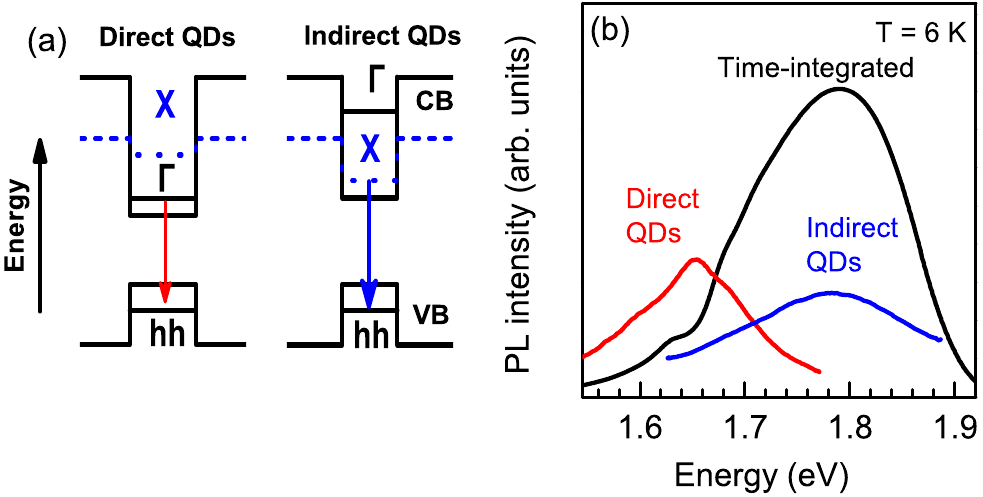}
\caption{\label{fig2} (a) Band diagrams of (In,Al)As/AlAs QDs with
direct and indirect band structures. Arrows mark the optical
transitions related to the decay of the ground state exciton.
(b) PL spectra of (In,Al)As/AlAs QDs measured for
nonresonant excitation: time-integrated (black line), time-resolved
for $t_{\text{delay}}=1$~ns and $t_{\text{gate}}=4$~ns (red) and
for $t_{\text{delay}}=1000$~ns and $t_{\text{gate}}=1500$~ns (blue).
$T=6$~K.}
\end{figure}

\begin{figure}[]
\includegraphics[width=\textwidth]{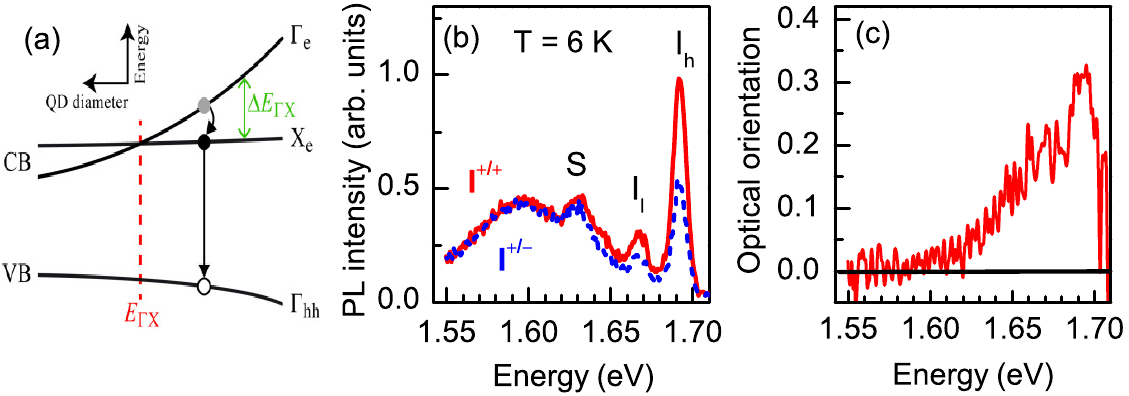}
\caption{\label{fig3} (a) Band alignment in QDs as function of
dot diameter for the valence (VB) and conduction (CB) bands. The energy
relaxation of a photo-excited electron from $\Gamma$ to X, and the
subsequent recombination are shown by the arrows. (b)
PL spectra of (In,Al)As/AlAs QDs measured in
${\sigma^+}$ and ${\sigma^-}$ polarization for ${\sigma^+}$
circularly polarized excitation, $E_{\text{exc}}=1.72$~eV, $T=6$~K.
(c) Optical orientation calculated from the data shown in panel
(b).}
\end{figure}

PL spectra of an (In,Al)As/AlAs QD ensemble measured for
nonresonant excitation are shown in Fig.~\ref{fig2}(b). The
time-integrated spectrum (black line) has its maximum at 1.79~eV and
extends from  1.5  to 1.9~eV having a full width at half maximum
(FWHM) of 190~meV. The large width of the emission band is due to
the dispersion of the QD parameters, since the exciton energy
depends on the QD size, shape, and composition~\cite{Shamirzaev78}.
The PL band is contributed by the emission of direct and indirect
QDs, which becomes evident from the time-resolved PL spectra. When
the spectrum is measured immediately after the laser pulse
($t_{\text{delay}}=1$~ns and $t_{\text{gate}}=4$~ns), the PL band
has the maximum at 1.65~eV and the FWHM of 120~meV (red line). For longer
delays ($t_{\text{delay}}=1000$~ns and $t_{\text{gate}}=1500$~ns),
the emission maximum shifts to 1.78~eV and broadens to 190~meV (blue
line), rather similar to the time-integrated PL spectrum.

We recently demonstrated that after photoexcitation in the
AlAs barriers, electrons and holes are captured into the QDs within
several picoseconds, and the capture probability does not depend on
the QD size and composition~\cite{ShamirzaevNT}. Therefore, all QDs
in the ensemble (direct and indirect ones) become equally populated
shortly after the excitation pulse. Thus, the exciton recombination
dynamics is fast for direct QDs emitting mainly in the spectral
range of $1.50${--}$1.74$~eV and slow for the indirect QDs emitting
in the range of $1.62${--}$1.90$~eV. The emissions of the direct
and indirect QDs overlap in the range of $1.62${--}$1.74$~eV.

\subsection{PL for selective excitation in zero magnetic field}

In order to excite only a fraction of QDs with indirect band gap we
used selective excitation within the inhomogeneously broadened PL
line. As a result, the PL band transforms into a spectrum with
rather narrow lines~\cite{Rautert99}. PL spectra measured for $\sigma^+$
excitation at $E_{\text{exc}}=1.72$~eV  using co- and
cross-polarized detection are shown in Fig.~\ref{fig3}(b).
As we showed in Ref.~\cite{Rautert99}, the lines
marked as $\text{I}_\text{l}$ and $\text{I}_\text{h}$ arise from
exciton recombination in the indirect QDs, while the line $S$
arises from a transition in QDs with $\Gamma$-$X$ mixing of the
electron states. Tuning the excitation energy allows us to
selectively excite different sub-ensembles of QDs.

\begin{figure}[]
\includegraphics[width=\textwidth]{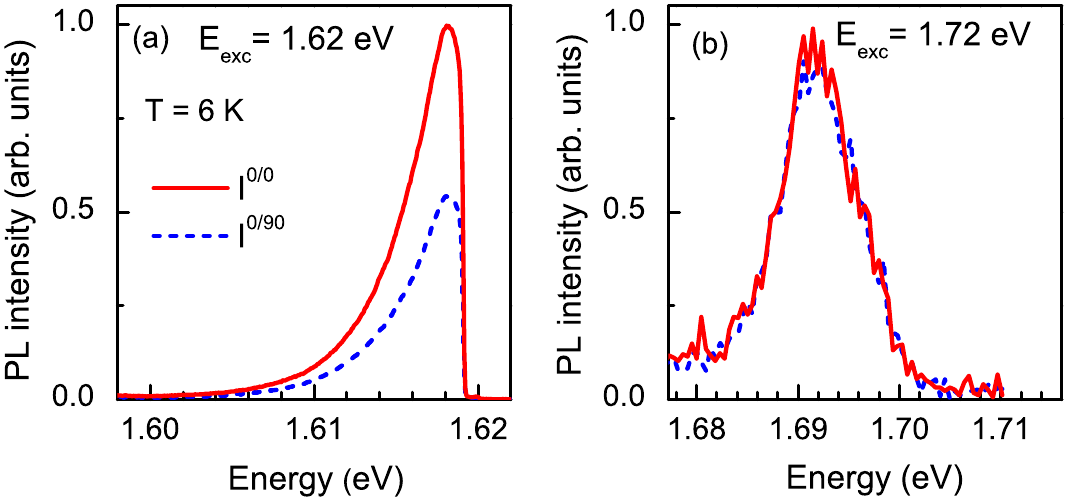}
\caption{\label{fig4} Linearly polarized PL spectra of
(In,Al)As/AlAs QDs measured for linearly  polarized excitation,
$T=6$~K. (a) Direct band gap QDs, $E_{\text{exc}}=1.62$~eV. (b)
Indirect band gap QDs, $E_{\text{exc}}=1.72$~eV.}
\end{figure}

Optical orientation across the PL spectrum for excitation at
$E_{\text{exc}}=1.72$~eV is shown in Fig.~\ref{fig3}(c). The PL in
the low-energy spectral region, which corresponds to exciton
recombination in direct band gap QDs, demonstrates almost zero optical
orientation. Contrary to that, the indirect QDs (high-energy
spectral region) demonstrate pronounced optical orientation that
reaches 0.3 (i.e. 30~$\%$) at the maximum of the $\text{I}_\text{h}$
line (at 1.695~eV). Linearly polarized emission for
linearly polarized excitation (optical alignment) is observed for
direct QDs, but is absent in indirect QDs, as shown in
Figs.~\ref{fig4}(a) and~\ref{fig4}(b), respectively.

These results are explained by the exciton fine structure. The
exciton is formed by a heavy hole with angular momentum
projection $j_z= 3/2$ and an electron with $s=1/2$ spin.
Accordingly, there are four exciton fine structure states. The two
bright exciton states are characterized by the angular momentum
projections $J_z=\pm 1$ onto the growth axis $z$ and the two dark
states have the projections $J_z=\pm 2$. The breaking of the axial
symmetry in direct QDs lifts the degeneracy of the bright exciton
states and mixes them so that the following states emerge:
$\ket{\text{X}}=\frac{1}{\sqrt{2}}(\ket{+1}+\ket{-1})$ and
$\ket{\text{Y}}=\frac{1}{i\sqrt{2}}(\ket{+1}-\ket{-1})$~\cite{Bayer}.
A circularly polarized photon excites a superposition of the
states $\ket{\text{X}}$ and $\ket{\text{Y}}$,  whose coherence is
rapidly lost, destroying the optical orientation of the
excitons~\cite{Paillard}. Linearly polarized photons, by
contrast, excite the pure states $\ket{\text{X}}$ and
$\ket{\text{Y}}$ of the bright exciton, so that the linear
polarization degree of the emission (optical alignment) is determined
by the ratio of the exciton spin decoherence time to the
exciton lifetime $\tau_\text{R}$. The high value of optical alignment for the direct QDs
of more than 30\% leads us to the conclusion
that the spin decoherence time exceeds the recombination one, which
is typical for direct band gap QDs~\cite{Paillard}. For
indirect band gap QDs, the anisotropic electron-hole
exchange interaction is negligible due to
the weak overlap of the wave functions of the
X-electron and the $\Gamma$-hole in momentum
space~\cite{Pikus,Bir74}, so that the pure exciton spin states $J_z = \pm
1$ provide circularly polarized PL~\cite{Rautert99}.

\subsection{Optical orientation in  longitudinal  magnetic field}
\label{sec:recovery}

\begin{figure}[]
\includegraphics[width=\textwidth]{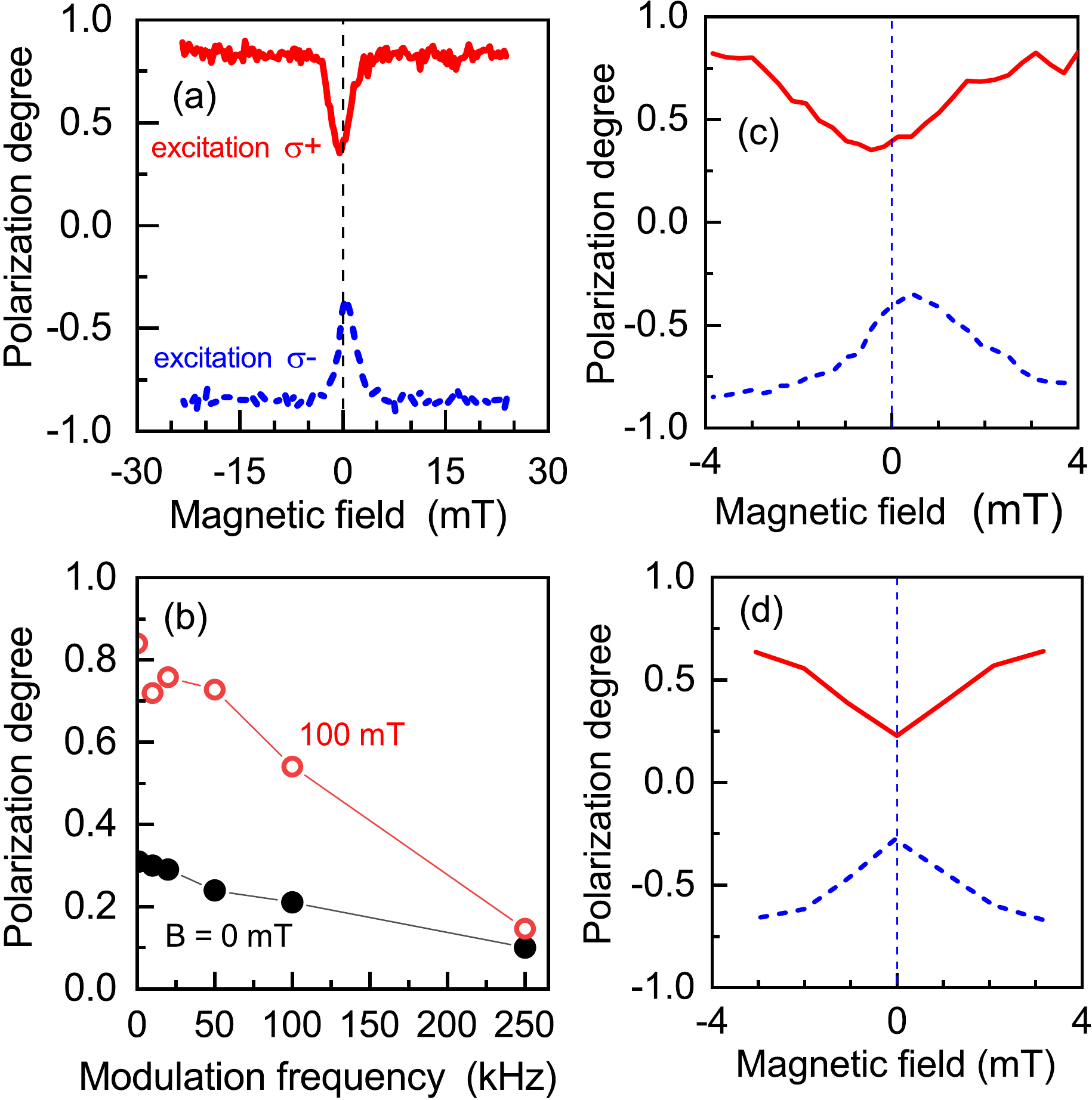}
\caption{\label{fig5} (a) Optical orientation in the indirect band
gap QDs at 1.695~eV for $E_{\text{exc}}=1.72$~eV at $T=6$~K.
Excitation with protocol (i): ${\sigma^+}$ (red solid line) and
${\sigma^-}$ (blue dashed line). (b) $\rho^0_\text{c}$ (black
circles) and $\rho_{\text{sat}}$ for $B=100$~mT (red circles) as
functions of the excitation polarization modulation frequency in
protocol (ii). (c) and (d) Details of the PL circular polarization
degree near zero external magnetic field for the protocols (i) and
(ii), respectively.}
\end{figure}

Optical orientation at the I$_h$ line maximum (1.695~eV) for
selective $cw$ excitation at $E_{\text{exc}}=1.72$~eV in protocol
(i) [see Fig.~\ref{fig1}(a)] as function of the longitudinal
magnetic field is shown in Fig.~\ref{fig5}(a). One can see that at
zero magnetic field the PL polarization degree $\rho^0_\text{c} =
0.31$. Already in  magnetic  fields of a few mT $\rho_\text{c}(B)$
demonstrates a strong change. The optical orientation gradually
increases with increasing magnetic field, and saturates at
$\rho_{\text{sat}}$=0.84, which is about 3 times larger than
$\rho^0_\text{c}$.

The shape of the polarization recovery curve (PRC) is described by a
Lorentz curve
$\rho_\text{c}(B)=\rho^0_\text{c}+(\rho_{\text{sat}}-\rho^0_\text{c})/
\left( 1 + \Delta^2_{\text{PRC}}/B^2 \right)$ with the half width at
half  maximum of $\Delta_{\text{PRC}}=1.8$~mT. We recently
demonstrated  that $\Delta_{\text{PRC}}$ arises from the electron spin precession in the local
fields created by the nuclear spin fluctuations~\cite{Merkulov},
which govern the electron spin dynamics in  magnetic fields $B
\thicksim \Delta_{\text{PRC}}$~\cite{Rautert99,Kuznetsova101,Smirnov125}.

We can estimate the anisotropic exchange interaction for the
indirect excitons as
${\delta_{\text{1}}}<\Delta_{\text{PRC}}\mu_\text{B}
g_{\text{e}}$~\cite{Rautert99}, where $\mu_{\text{B}}$ is the Bohr
magneton and $g_\text{e}$ is the electron $g$-factor. Using
$g_{\text{e}}=2$~\cite{Debus,Ivanov97,Ivanov104}, we get
${\delta_{\text{1}}}<0.2~\mu$eV, which is indeed several orders of
magnitude smaller than the ${\delta_{\text{1}}}$ of several hundreds
of $\mu$eV observed in direct band gap (In,Al)As/AlAs
QDs~\cite{Rautert100}.

The increase of the PL polarization degree in a longitudinal magnetic
field by a factor of about 3 (from $\rho^0_\text{c}=0.31$ to
$\rho_{\mathrm{sat}}=0.84$) indicates that the electron spin
relaxation time $T_1$ is longer than the indirect exciton
recombination time. Indeed, the electron spin in a QD undergoes
Larmor precession around the effective frozen nuclear field,
$\mathbf{B}_N$, induced by the nuclear spin fluctuations. The
photogenerated spin-oriented electrons lose 2/3 of their spin
polarization during the time
$T_2^*\sim\hbar/(g_e\mu_B\Delta_{\text{PRC}})$, since $\mathbf{B}_N$
has no preferential orientation and its direction varies from dot to
dot in a QD ensemble. The rest 1/3 of the electron spin polarization
is stabilized via the interaction with the nuclear spins pointing along
the orientation direction, i.e. the
$z$-axis~\cite{Rautert99,Kuznetsova101}. The deviation of
$\rho_{\mathrm{sat}}$ from unity is the result of the loss of the
electron and hole spin polarizations during energy
relaxation via the transition from the $\Gamma$-$\Gamma$ exciton to
the $\Gamma$-X exciton ~\cite{Kuznetsova101}. Note that the loss
of electron spin polarization alone cannot describe
$\rho_{\text{sat}}<1$ because the hole spin uniquely defines the
emitted photon polarization.

\begin{figure}[]
\includegraphics[width=\textwidth]{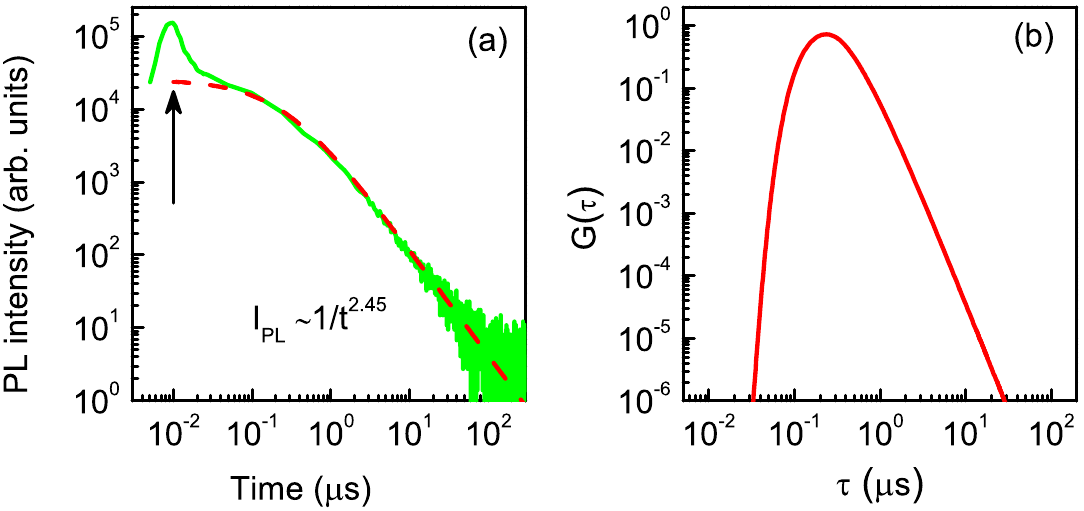}
\caption{\label{fig6} (a) PL  dynamics measured at the detection
energy of 1.695~eV. The excitation pulse with the photon energy of
3.49~eV ends at $t=10$~ns (marked by the vertical arrow). The dashed curve
is a fit with the parameters given in the text. (b) Normalized exciton
lifetime distribution function $G(\tau$) corresponding to the QD
subensemble emitting at the energy of 1.695~eV, obtained by
fitting the PL dynamics.}
\end{figure}

Taking into account the exciton lifetime in the QDs, we can estimate the
lower boundary for the spin relaxation time $T_1$ of electrons in
indirect band gap QDs in a longitudinal  magnetic field, which
eliminates the effect of the nuclear field on the electron spin
dynamics. In order to determine  the typical exciton lifetime in
QDs, we measured the PL dynamics at the detection energy of 1.695~eV
for nonresonant excitation, as shown in Fig.~\ref{fig6}(a). The
PL dynamics are plotted on a double-logarithmic scale, which
is convenient to cover the wide range of scanned times and PL intensities.

The recombination dynamics demonstrate two distinctive stages: (i)
a fast PL decay immediately after the excitation pulse corresponding
to recombination in the direct band gap QDs, since direct
and indirect QDs are coexisting in this spectral region,
see Fig.~\ref{fig2}; (ii) a further PL decay that can be described
by a power-law function $I(t)\sim (1/t)^\alpha$, as shown in our
previous studies~\cite{Shamirzaev78,Shamirzaev84,Ivanov97}. Such a
dynamics result from the superposition of multiple monoexponential
decays with different times varying with size, shape, and
composition of the indirect band gap QDs. It can be described by the
following equation~\cite{Nikolaev,Shamirzaev84}:
\begin{equation} \label{eq1}
I(t) = \int_0^{\infty} G(\tau) \exp \left( -\frac{t}{\tau} \right )
d\tau,
\end{equation}
where $G(\tau)$ is the distribution function of the exciton
recombination times $\tau$. It has the rather simple
form~\cite{Shamirzaev84}:
\begin{equation} \label{eq2}
G(\tau) = \frac{C}{\tau^{\gamma}}
\textnormal{exp}{\left(-\frac{\tau_{0}}{\tau}\right)}.
\end{equation}
Here $C$ is a constant and $\tau_0$ characterizes the maximum of the
distribution of the exciton lifetimes. The parameter $\gamma$ can
be extracted directly from the power-law decay (1/$t$)$^{\gamma-1}$
presented in Fig.~\ref{fig6}(a). Using the model approach
suggested in our recent study~\cite{Shamirzaev84} we have obtained
this distribution function via fitting the recombination dynamics
in Fig.~\ref{fig6}(a), see the dashed line. The fit parameters are
$\gamma =3.45$  and $\tau_{0} =0.25~\mu$s. The distribution $G(\tau)$ is shown
in Fig.~\ref{fig6}(b). The typical recombination time, $\tau_{0}$,
for excitons in the QD subensemble emitting at 1.695~eV equals to
0.25~$\mu$s. Thus, the typical spin relaxation time $T_1$ of
electrons in indirect band gap QDs in a longitudinal  magnetic field
is longer than $\tau_{0}=0.25~\mu$s.

\subsection{Effect of excitation-detection protocol on the optical
  orientation}
\label{sec:protocol}

Continuous excitation of localized electrons by circularly
polarized light [as we apply in the protocol (i)] can lead via
the Knight field to polarization of the nuclear spins, i.e. to
dynamic nuclear polarization
(DNP)~\cite{Dyakonov,OO_book,Urbaszek}. The nuclear polarization
degree is determined by the ratio between the spin
transfer rate from electrons to the nuclei and the nuclear spin relaxation
rate~\cite{Dyakonov}. The nuclear spin relaxation times in
A$_3$B$_5$ semiconductors can reach several seconds, and the
Overhauser field of the polarized nuclei acting on the electrons
can reach several Tesla~\cite{OO_book}.

In our case, the DNP manifests itself as a shift of the minimum of the
PRC by 0.5~mT from the zero field position, see Fig.~\ref{fig5}(c).
A change of the excitation polarization (from $\sigma^+$ to
$\sigma^-$) results in a change of the shift direction to the
opposite one. Note that the value of the DNP-induced Overhauser
field in our indirect band gap QDs is smaller than the typical one
(about $10-20$~mT) observed for direct band gap (In,Ga)As QDs at
comparable excitation conditions~\cite{Kuznetsova89}. The relatively
weak DNP-induced Overhauser field in indirect band gap (In,Al)As QDs
originates from two specific features of this system: (i) The long
exciton lifetime reduces the rate of spin transfer from the electrons
to the nuclei. Indeed, the number of electrons that have the possibility to
transfer spin polarization to the nuclei in direct band gap systems,
which have a typical exciton lifetime of a nanosecond, is about
10$^9$ per second. On the other hand, in systems with indirect band gap,
where the exciton lifetime is about a microsecond, this number
decreases by several orders of magnitude. (ii) As recently
shown, the hyperfine interaction constant for an electron in the X
valley of (In,Al)As QDs with the As nuclei is about two times, and with
the In and Al nuclei about two orders of magnitude smaller than for
an electron in the $\Gamma$ valley~\cite{Kuznetsova101}. Thus the
Overhauser field induced by polarized nuclei in indirect QDs is
several times smaller than the one in direct band gap QDs even
at a similar nuclear spin polarization degree.

A common technique for DNP suppression during optical orientation is
the modulation of the helicity of the exciting light~\cite{OO_book}.
We used this technique [excitation corresponding to the protocol
(ii)] for the measurement of PRCs. Figure~\ref{fig5}(d) demonstrates
the absence of a shift of the PRCs at $f_{\text{m}}=10$~kHz, which
evidences the DNP suppression.

However, a strong difference in optical orientation occurs when
using the excitation protocol (ii) compared with protocol (i).
$\rho^0_\text{c}$ and $\rho_{\mathrm{sat}}$ for $B=100$~mT are shown in Fig.~\ref{fig5}(b) as
functions of the modulation frequency of the
excitation polarization. One can see
that $\rho^0_\text{c}$ decreases with increasing $f_{\text{m}}$ from
0.31 at $f_{\text{m}}=0$ down to 0.10 at $f_{\text{m}}=250$~kHz.
$\rho_{\mathrm{sat}}$, equal to 0.84 at $f_{\text{m}}=0$,
decreases down to 0.15 at $f_{\text{m}}=250$~kHz.

\begin{figure}[]
\includegraphics[width=0.5\textwidth]{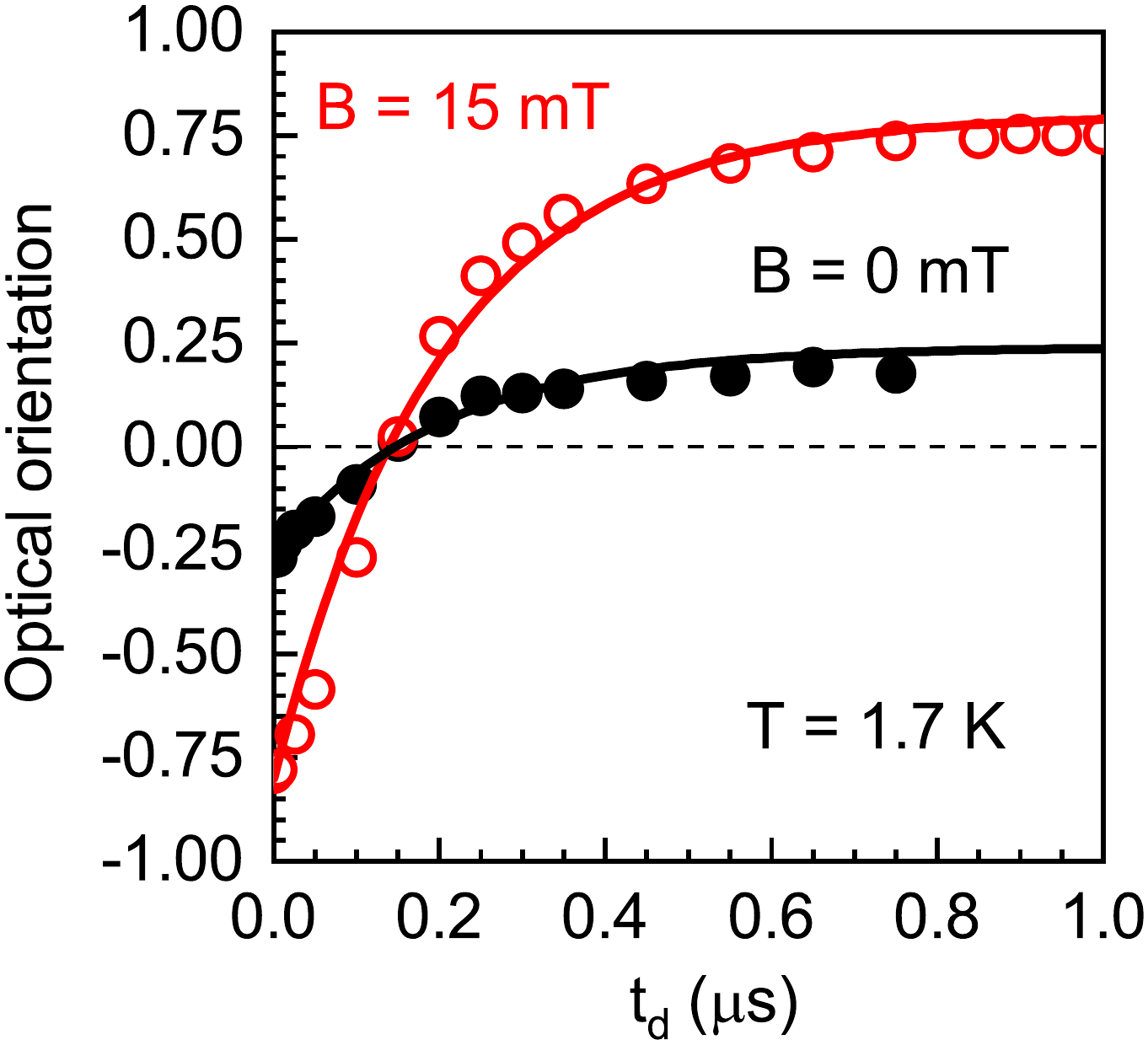}
\caption{\label{fig7} Optical orientation $\rho^0_\text{c}$ (filled
black circles) and $\rho_{\mathrm{sat}}$ for $B = 15$~mT (open red
circles) as functions of the delay time $t_{\text{d}}$, measured at
$t_{\text{ex}} = 1~\mu$s and $t_{\text{w}} = 0.05~\mu$s. $T =
1.7$~K. $t_{\text{d}}=0$ corresponds to a change of excitation
polarization from ${\sigma^-}$ to ${\sigma^+}$. The solid lines show
fits after Eq.~\eqref{eq5} with the parameters given in the text.}
\end{figure}

In order to understand these results, we use the protocol (iii) [see
Fig.~\ref{fig1}(c)] with different excitation times,
$t_{\text{ex}}$, delay times, $t_{\text{d}}$, and measurement time
windows, $t_{\text{w}}$. The dependences of
$\rho^0_\text{c}$($t_{\text{d}}$) and
$\rho_{\mathrm{sat}}$($t_{\text{d}}$) for $B = 15$~mT, measured at
$t_{\text{ex}} = 1~\mu$s and $t_{\text{w}} = 0.05~\mu$s, are shown in
Fig.~\ref{fig7}. One can see that both $\rho^0_\text{c}$ and
$\rho_{\mathrm{sat}}$ at zero delay time, corresponding to the
change of excitation polarization from ${\sigma^-}$ to
${\sigma^+}$, surprisingly are negative (i.e., are dominated by the
counter-polarized $I^{+/-}$ PL component) and equal to $-0.75$
$(-0.25)$ for $\rho_{\mathrm{sat}}$ ($\rho^0_\text{c}$). When the
delay time increases, the optical orientation decreases to zero
at $t_{\text{d}} = 0.15~\mu$s. A further increase in $t_{\text{d}}$
changes the polarization to positive values (dominated by the
co-polarized $I^{+/+}$ PL component) and the polarization degree
increases to $+0.75$ $(+0.25)$ for $\rho_{\mathrm{sat}}$
($\rho^0_\text{c}$) at $t_{\text{d}} = 1~\mu$s.

The change of the excitation polarization upon modulation of the
exciting light in protocols (ii) and (iii) occurs on much shorter
timescales than the indirect exciton lifetime. This results in a
situation, where after the change of excitation polarization, a
fraction of the QDs is still occupied by excitons created in the
previous half period of excitation with the corresponding direction
of spin polarization, while the other fraction of them begins to
become occupied with excitons of opposite spin polarization. Both
types of excitons recombine simultaneously emitting oppositely
polarized photons. The ratio of oppositely polarized exciton
concentrations changes with time. Using continuous detection, we
measure the integral from all of these processes, which varies with
the modulation frequency of the excitation polarization (see
Fig.~\ref{fig5}(b)).

We describe the dependence of the optical orientation on the delay after
changing the excitation polarization as follows:
\begin{equation}
\label{eq5}\rho_{\text{c}}(t) = \frac{[I^{+/+}(t)+I^{-/+}(t)] -
[I^{+/-}(t)+I^{-/-}(t)]}{[I^{+/+}(t)+I^{-/+}(t)] + [I^{+/-}(t)
+I^{-/-}(t)]}=\rho_{\text{c}}^{\text{e}}[1-2\exp(-t/\tau)],
\end{equation}
where $\rho_{\text{c}}^{\text{e}}$ is the circular polarization degree
at the end of the excitation period for $t = t_{\text{ex}}$. One can
see that the experimental data in Fig.~\ref{fig7} can be well fitted
by Eq.~{\eqref{eq5}} with the exciton recombination time $\tau =
0.21~\mu$s, which is in reasonable agreement with the $\tau_0=0.25~\mu$s obtained
from the PL dynamics measurements in Sec.~\ref{sec:recovery}.

\begin{figure}[]
\includegraphics[width=0.6\textwidth]{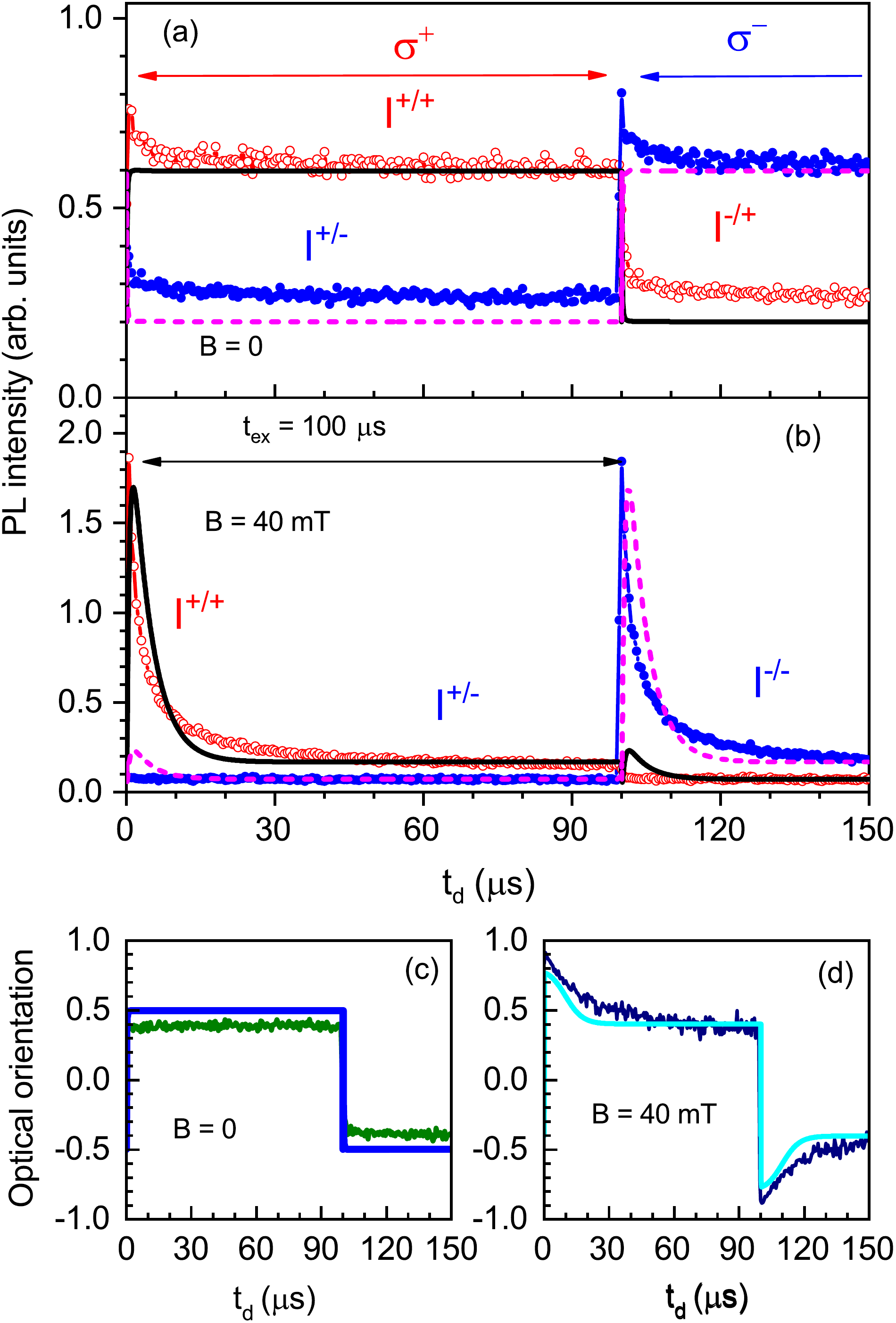}
\caption{\label{fig8} Intensities of the PL components labeled in
the figure as functions of the delay time $t_{\text{d}}$ for (a) $B
= 0$~mT, and (b) $B = 40$~mT, measured at $t_{\text{ex}} =
100~\mu$s and $t_{\text{w}} = 2~\mu$s. $T = 1.7$~K. The optical orientation
corresponding to these functions are presented in panels (c) for $B
= 0$~mT and (d) for $B = 40$~mT. The theoretical simulations with the
parameters given in the text are shown by the solid and dashed lines for
the corresponding external magnetic fields.}
\end{figure}

In order to study the spin dynamics at time scales that
strongly exceed the exciton lifetimes, we measured the intensities
of the co- ($I^{+/+}$) and counter- ($I^{+/-}$) polarized PL components
for $t_{\text{ex}} = 100~\mu$s and $t_{\text{w}} = 2~\mu$s. The
results of these measurements at zero magnetic field and in an
magnetic field of 40~mT (which corresponds to $\rho_{\text{sat}}$)
are shown in Fig.~\ref{fig8} as functions of $t_{\text{d}}$. At zero
magnetic field, both co- and counter-polarized PL component
intensities after a short transient process, which follows the
change of the excitation polarization, have identical temporal
dependences [see Fig.~\ref{fig8}(a)], resulting in a constant
polarization [see Fig.~\ref{fig8}(c)]. However, these dependences
change drastically in magnetic field. The intensity of the
counter-polarized PL component does not depend on delay time, while
the co-polarized PL component increases strongly with the change of
excitation polarization, namely by an order of magnitude in
Fig.~\ref{fig8}(b), and then decays with increasing $t_{\text{d}}$.
Thus, the PL polarization degree follows the intensity of the
co-polarized component [see Fig.~\ref{fig8}(d)].

\begin{figure}[]
\includegraphics[width=0.5\textwidth]{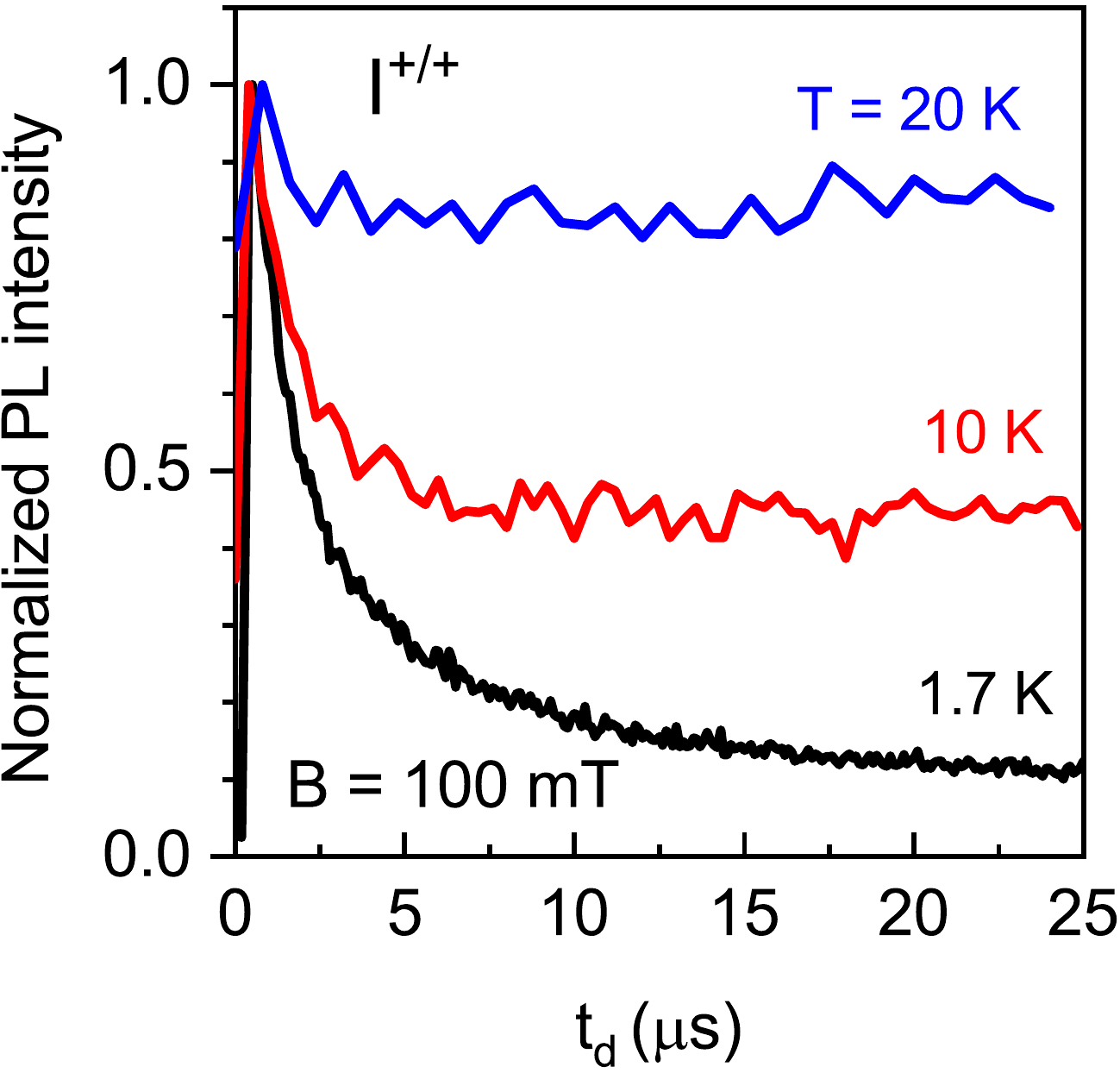}
\caption{\label{fig9} Intensity of the co-polarized ($I^{+/+}$)
PL component as function of the delay time $t_{\text{d}}$
for $B = 100$~mT, measured at $t_{\text{ex}} = 25~\mu$s,
$t_{\text{w}} = 2~\mu$s, and different temperatures.}
\end{figure}

Finally, Fig.~\ref{fig9} shows that at higher temperatures the
decay of the co-polarized PL ($I^{+/+}$) with the delay time $t_d$
becomes weaker.

In conclusion of this section, we summarize the most important
experimental findings. The magnetic field dependence of the
optical orientation in (In,Al)As/AlAs QDs is changing strongly with the
measurement protocol:

(i) The optical orientation depends on the modulation frequency
of the excitation polarization.

(ii) Measurement of the $\sigma^+$ and $\sigma^-$ polarized
PL components in a short time window $t_{\text{w}}$ with delay
$t_{\text{d}}$ after changing the excitation polarization (from
$\sigma^+$ to $\sigma^-$ and vice versa) allows us to reveal the lifetime
and other features of the exciton spin dynamics in indirect QDs.

(iii) At zero magnetic field, both the co- and counter-polarized PL
component intensities have identical time dependences, while in a
magnetic field with a strength exceeding the fluctuations of the
nuclear field, the intensity of the co-polarized PL component
increases strongly with changing the excitation polarization and
then decays with increasing $t_{\text{d}}$. However, the intensity
of the counter-polarized PL component does not depend on the delay
time.

(iv) The decrease of the co-polarized PL component intensity with
delay time $t_{\text{d}}$ in magnetic field disappears with
increasing temperature.

\section{Discussion}
\label{sec:4}

The most surprising experimental result is shown in Fig.~\ref{fig8}.
Figure~\ref{fig8}(b) evidences that after switching the excitation
polarization, the intensity of the co-polarized emission, $I^{+/+}$ or
$I^{-/-}$, changes strongly at $t_d \sim \addDima{10} \, \mu s$. This happens
in a magnetic field of $40$~mT, while at zero magnetic field there
are no such changes, as demonstrated in Fig.~\ref{fig8}(a).

In some systems, the decrease of the PL intensity may be related to
the suppression of the mixing of dark with bright
excitons~\cite{Rho,Moire} after DNP. However, in our case, we have shown in
Sec.~\ref{sec:recovery} that the splitting between the bright and dark
excitons is small, so that at least half of the excitons created by
quasi-resonant excitation are bright. In this case the suppression
of the mixing between bright and dark excitons cannot explain the
decrease of the PL intensity by an order of magnitude.

\begin{figure}
\includegraphics[width=0.6\textwidth]{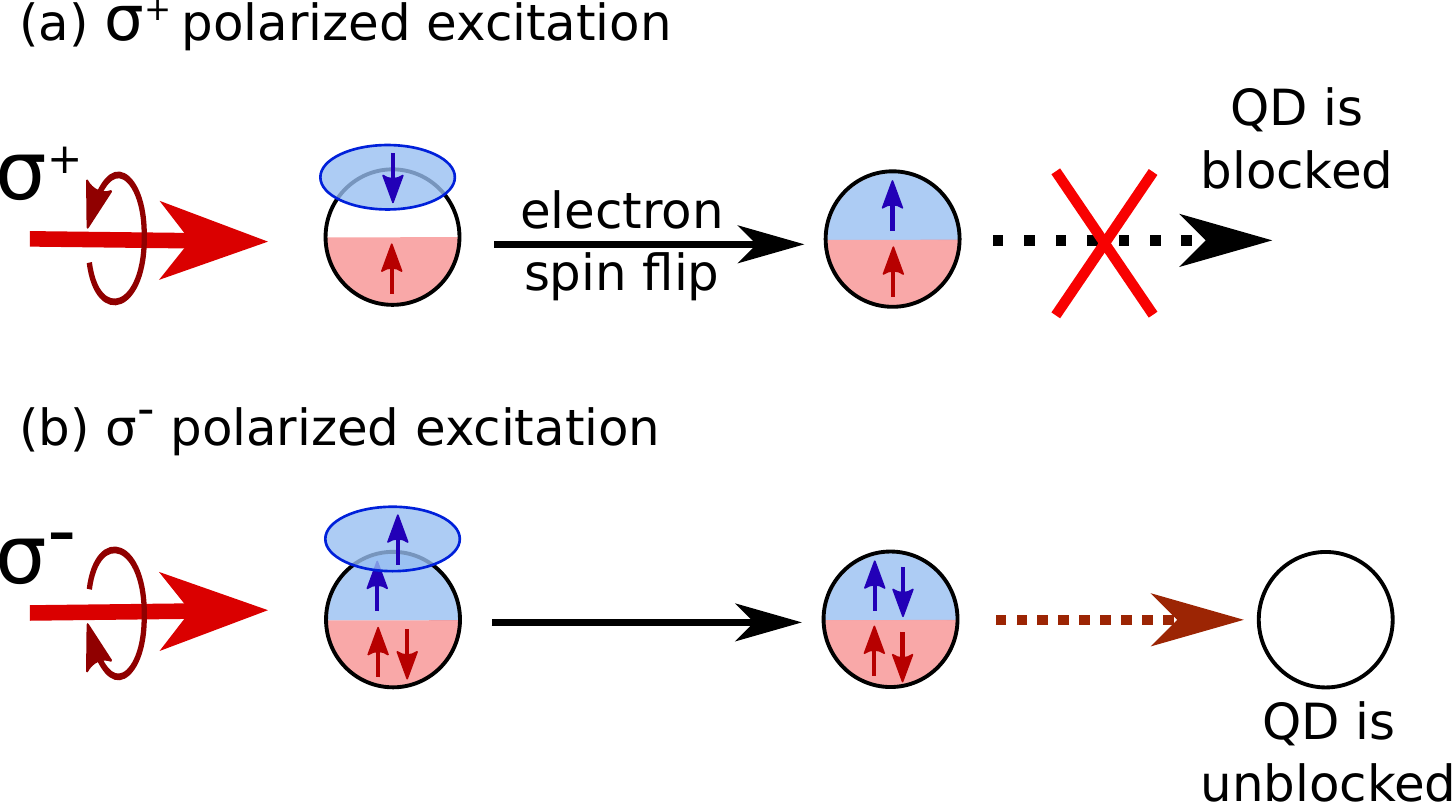}
\caption{\label{figM} Blocking of QD due to dark exciton creation
(a) and its unblocking after change of the excitation
polarization (b).}
\end{figure}

We suggest that the observed effect is related to blockade of the
QDs by dark excitons. Let us qualitatively describe the mechanism of
decrease of the co-polarized PL intensity with time. In a
strong enough longitudinal magnetic field, the nuclei-induced mixing
between the bright and dark excitons is negligible. In this case,
$\sigma^+$ excitation creates mostly bright excitons, however, due to
electron spin relaxation, dark excitons can be also created.
These excitons have long lifetimes that are controlled by the hole spin-flip
rate $\gamma_h$. While each QD can occupy only a single one, dark excitons can accumulate in the ensemble
and occupy a significant fraction of the QDs, leading to suppression of the PL intensity
[Fig.~\ref{figM}(a)].

When the excitation polarization is switched from $\sigma^{+}$ to
$\sigma^{-}$, the possibility appears for QDs to capture a second
photon and form a biexciton. Fast biexciton
recombination returns the blocked QDs to an optically active state, so
that the PL intensity recovers [Fig.~\ref{figM}(b)]. The
Pauli exclusion principle forbids the biexciton formation for the
initial excitation polarization.

This phenomenon occurs when the applied magnetic fields are strong enough.
In zero magnetic field, the bright and dark excitons become in effect
mixed by the random nuclear field, so that all of the four exciton
types can recombine radiatively quickly. As a result, the PL
intensity does not change strongly with time as shown in
Fig.~\ref{fig8}(a).

With increasing temperature, the electron spin relaxation
accelerates, so that the effect of the PL intensity decrease with time
disappears, in agreement with Fig.~\ref{fig9}.

\subsection{Theory of QD blockade}
\label{sec:theory}

In this section we give a detailed model of the QD blockade by the dark excitons.

\begin{figure}[]
\includegraphics[width=0.6\textwidth]{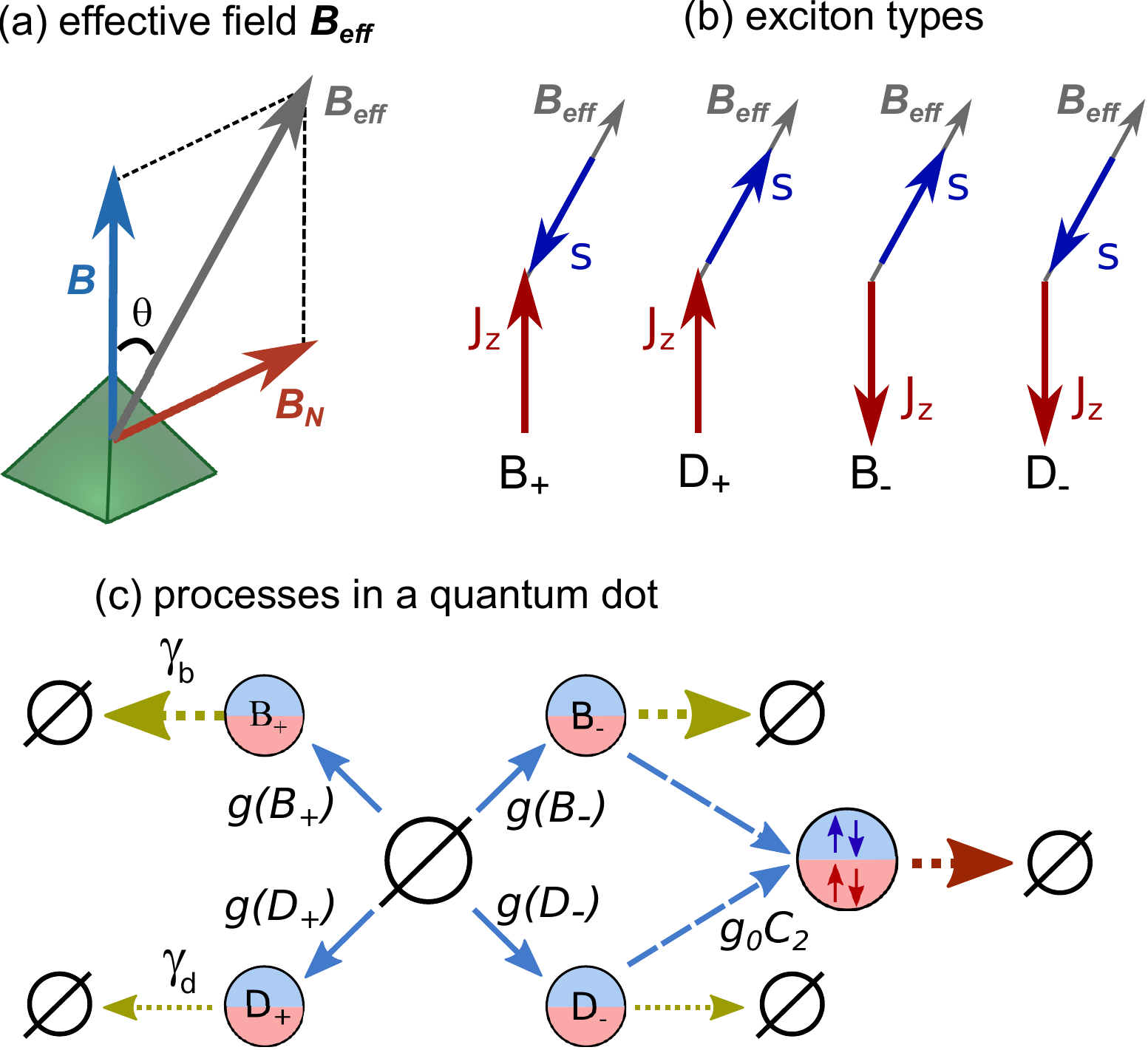}
\caption{\label{figMod} (a) Total effective magnetic field ${\bm
B}_{\text{eff}}$ scting on an electron, which is composed of
the random nuclear field ${\bm B}_N$ and the external field ${\bm
B}$ parallel to the growth axis $z$.  (b) The four exciton
eigenstates with the electron spin (blue arrow) along or opposite
to the effective magnetic field (grey arrow) and the hole spin (red
arrow) along or opposite to the $z$ axis. (c) Possible transitions
between QD states during $\sigma^+$ excitation, $\bm\varnothing$
denotes the empty QD state.}
\end{figure}

For quasi-resonant excitation by $\sigma^+$ ($\sigma^-$) polarized
light, the bright excitons are created in the QDs with a spin-up
(-down) heavy hole and a spin-down (-up) electron. We assume that
shortly after excitation, electron and hole can flip their
spins with the probabilities $f_e$ and $f_h$, respectively, during
the exciton relaxation. As a result, the occupancies of the
electron spin-up and spin-down states are $f_e$ ($1-f_e$) and
$1-f_e$ ($f_e$), and similarly for the hole spin states. The spin
flips may be related to the electron-hole exchange interaction in
the direct momentum exciton state or to the electron-phonon
interaction during the electron energy relaxation. \addDima{For the ground state, we} describe the electron spin dynamics by the precession
in the effective magnetic field ${\bm B}_{\text{eff}}$ composed of
the external field ${\bm B}$ and the random nuclear field ${\bm
B}_N$~\cite{Smirnov125,Shamirzaev104a}, see Fig.~\ref{figMod}(a). We
assume the nuclear field to be quasi-static and Gaussian
distributed as
$\propto \exp(-B_N^2/\Delta_B^2)$~\cite{Merkulov,Glazov,myUFN},
neglecting the anisotropy of the hyperfine
interaction~\cite{Kuznetsova101} and the intervalley hyperfine
interaction~\cite{Avdeev}. Due to the light polarization
modulation, a DNP is absent~\cite{Zhukov}, so we neglect it as well as
the nuclear spin dynamics, which in principle can take place on a
submillisecond time scale~\cite{CSM}.

We assume the electron spin precession \addDima{(the typical period is of the order of 10~ns)} to be faster than the
exciton recombination, so the direction of ${\bm B}_{\text{eff}}$
defines the appropriate quantization axis for the electron spin.  We denote
the exciton states with the electron spin along or opposite to the
direction of ${\bm B}_{\text{eff}}$ and the hole spin-up or -down as $B_\pm$
and $D_\pm$, respectively, see Fig.~\ref{figMod}(b). In a strong
longitudinal magnetic field, $B_z\gg B_N$, the effective magnetic
field is almost parallel to the $z$ axis, so $B_\pm$ and $D_\pm$ are
quasi-bright and quasi-dark states, respectively.

As described above, we consider the generation of all four exciton
states starting from an empty QD. Under $\sigma^+$ excitation the corresponding
generation rates have the form
\begin{equation} \label{gBp}
g(B_+) = g_0 \left[  (1 - f_h)(1-f_e) \frac{1+\cos\theta}{2} +
(1 - f_h)f_e \frac{1-\cos\theta}{2} \right],
\end{equation}
\begin{equation} \label{gDp}
g(D_+) = g_0 \left[ (1 - f_h)f_e \frac{1+\cos\theta}{2} +
(1 - f_h)(1-f_e) \frac{1-\cos\theta}{2} \right],
\end{equation}
\begin{equation} \label{gBm}
g(B_-) = g_0 \left[ f_h f_e \frac{1+\cos\theta}{2} \right.\left. +  f_h(1-f_e) \frac{1-\cos\theta}{2} \right],
\end{equation}
\begin{equation} \label{gDm}
g(D_-) = g_0 \left[ f_h (1 -f_e) \frac{1+\cos\theta}{2}  +  f_h f_e \frac{1-\cos\theta}{2} \right].
\end{equation}
Here $g_0$ is the pumping rate and $\theta$ is the angle between
$\bm B_{\text{eff}}$ and the $z$ axis, see
Fig.~\ref{figMod}(a). For $\sigma^-$ excitation, the subscripts of
$D_\pm$ and $B_\pm$ should be flipped.

We take into account the radiative recombination of the bright excitons
with the rate $\gamma_r=1/\tau$. In analogy with the generation
rates, we find the rates of $B_\pm$ and $D_\pm$ exciton
recombination as given by
\begin{equation}
\gamma_b = \gamma_r \frac{1 + \cos\theta}{2}, \quad
\gamma_d = \gamma_r \frac{1 - \cos\theta}{2},
\end{equation}
respectively, see Fig.~\ref{figMod}(c). We also consider the
possibility of a hole spin flip with the rate $\gamma_h$.

In addition to that, we allow for biexciton formation in a QD,
as shown in Fig.~\ref{figMod}(c). We assume that under $\sigma^\pm$
excitation it can be formed from the $B_\mp$ and $D_\mp$ excitons only
with the rate $g_0C_2$ due to the Pauli exclusion principle for the
heavy hole spin. The biexciton recombination rate is assumed to be
higher than all other recombination rates for simplicity, which, however,
hardly affects the results. The biexciton resonance in
(In,Al)As/AlAs QDs is detuned from the exciton
one~\cite{Shamirzaev78}, therefore, biexciton PL is not detected.
However, after biexciton recombination, the QD can be excited
once again, so that it becomes optically active. Thus, the role of
biexciton generation is to facilitate dark exciton recombination
and to unblock the QDs after change of the excitation
polarization, see Fig.~\ref{figM}.

The kinetic equations for this model read
\begin{equation}\label{kinB}
\frac{\d n(B_\pm)}{\d t} = g(B_\pm) n(\varnothing) + \frac{\gamma_h}{2}\left[n(D_{\mp}) - n(B_\pm)\right]
- \gamma_b n(B_\pm) -  g_0C_2 \frac{1\mp\sigma}{2} n(B_\pm),
\end{equation}
\begin{equation}\label{kinD}
\frac{\d n(D_\pm)}{\d t} = g(D_\pm) n(\varnothing) + \frac{\gamma_h}{2}\left[n(B_{\mp}) - n(D_\pm)\right]
- \gamma_d n(D_\pm) - g_0C_2 \frac{1\mp\sigma}{2} n(D_\pm).
\end{equation}
Here $n(B_\pm)$ and $n(D_\pm)$ are the occupancies of the
corresponding excitonic states, and $n(\varnothing)$ is the
probability for a QD to be unoccupied. Due to the assumption of a
fast biexciton recombination rate, it is given by $n(\varnothing) =
1 - n(B_+) - n(B_-) - n(D_+) - n(D_-)$. $\sigma=\pm$ denotes
$\sigma^\pm$ polarization of light. The processes described by
these kinetic equation are shown in Fig.~\ref{figMod}(c) for
$\sigma=+1$.

The intensities of $\sigma^\pm$ PL are given by
\begin{equation}\label{inte}
  I^\pm \propto \gamma_r \left\langle   \frac{1 + \cos\theta}{2} n(B_{\pm}) +  \frac{1 - \cos\theta}{2} n(D_{\pm})   \right\rangle,
\end{equation}
where the angular brackets denote the averaging over the nuclear
field distribution.

\subsection{Modeling of experimental results}

To describe the experimental data by this model, we numerically calculate the PL intensity and polarization
as functions of the delay time $t_{\text{d}}$ for $B = 0$~mT and $B
= 40$~mT. The averaging is performed over 100 random
realizations of ${\bm B}_N$. The comparison between the theoretical
and experimental results is shown in Fig.~\ref{fig8}. The
best fit was obtained using the parameters $g_0
= 2.2\, \mu{\rm  s}^{-1}$, $\gamma_r = 13\, \mu {\rm s}^{-1}$,
$\gamma_{h} = 0.035\, \mu {\rm s}^{-1}$, $C_2 = 0.5$, and $\Delta_B
= 0.33 \, {\rm mT}$. Note that the last parameter is not determined
reliably, it should be smaller than $40$~mT. The spin-flip
probabilities are $f_h = 0.3$ and $f_e = 0.25$. One can see, that
the agreement between theory and experiment is good.

The most reliably determined parameter is $\gamma_h$, because it
describes the decrease of the intensity with time in a strong magnetic
field. We also note that the obtained value of $\gamma_r$ \addDima{agrees in the order of magnitude} with the radiative lifetimes determined independently
in Secs.~\ref{sec:recovery} and~\ref{sec:protocol}. A specific
feature of our model is that even in strong magnetic fields, the
hyperfine interaction plays a role, because it can cause the
recombination rate of the quasi-dark excitons to be comparable to the slow
hole spin relaxation rate. In principle, there can be also
nonradiative recombination and electron spin flips, which have the
same effect. Another feature of the model is the absence of the electron hole
exchange interaction, which supports the previous
suggestion that it is weak~\cite{Kuznetsova101,Smirnov125}.

\section{Conclusion}
\label{sec:conclusions}

The exciton recombination and spin dynamics in indirect band gap
(In,Al)As/AlAs QDs with type-I band alignment have been studied
in a longitudinal
magnetic field by means of optical orientation.
We have demonstrated that the commonly used technique of measuring the optical
orientation based on modulation of the excitation
polarization with continuous-wave detection gives ambiguous results,
which depend on the modulation frequency due to long exciton
recombination and spin relaxation times in this system. A technique
based on measuring with a delay after change of the
excitation polarization has been proposed for overcoming this
problem. A QD blockade by dark excitons has been
revealed using this technique. The experimental findings have been
quantitatively described by a theoretical model accounting for the
population dynamics of the bright and dark exciton states as well as
biexciton formation in the QDs.

\vspace{6pt} 



\authorcontributions{
  Conceptualization and methodology, T.S.S.; theoretical investigation, D.S.S and A.V.S; investigation S.V.N.; writing software and optimization of the calculation program, A.V.S; software for the experiment, D.K.; funding acquisition, M.B., T.S.S., Yu.G.K. and D.S.S.; writing---original draft preparation T.S.S. and D.S.S.; writing---review and editing D.R.Y., T.S.S., D.S.S., S.V.N., Yu.G.K. and M.B. All authors have read and agreed to the published version of the manuscript.
}

\funding{The work of D.K., D.R.Y. and M.B. was supported by the Deutsche Forschungsgemeinschaft via the project No. 409810106. D.S.S. thanks the Foundation for the Advancement of Theoretical Physics and Mathematics ``BASIS.'' The development of the analytical theoretical model of QD blockade by D.S.S. was supported by the Russian Science Foundation, grant No. 21-72-10035. All experimental activities by T.S.S. including sample growth, microscopy, investigation of the energy level spectrum, magneto-optical properties as well as exciton recombination and spin dynamics were supported by a grant of the Russian Science Foundation (No. 22-12-00022). Numeric calculations of luminescence polarization and intensity by A.V.S. were supported by a grant of the Russian Science Foundation (No. 22-12-00125).}

\acknowledgments{We thank \href{http://www.ioffe.ru/coherent/index.html/Coherent/Ivchenko.html}{E. L.
Ivchenko} and \href{https://orcid.org/0000-0002-0454-342X}{M. O.
Nestoklon} for fruitful discussions.}

\conflictsofinterest{The authors declare no conflict of interest. The funders had no role in the design of the study; in the collection, analyses, or interpretation of data; in the writing of the manuscript; or in the decision to publish the~results.} 

\begin{adjustwidth}{-\extralength}{0cm}

\reftitle{References}

\PublishersNote{}
\end{adjustwidth}
\end{document}